\documentclass[aps,preprint,showpacs]{revtex4}
\usepackage{graphicx}
\usepackage{epsfig}
\begin{document}

\title{Geometrical constraints in a gene network model
and pattern formation}
\author{L. Diambra\footnote{Electronic address: diambra@if.sc.usp.br (corresponding author)} and
Luciano da Fontoura Costa\footnote{Electronic address:
luciano@if.sc.usp.br}} \affiliation{Institute of Physics at S\~ao
Carlos, University of S\~ ao Paulo. \\
Caixa Postal 369, cep 13560-970, S\~{a}o Carlos, SP, Brazil. \\
Phone +55 16 3373 9858, FAX +55 16 3373 9879}

\begin{abstract} A fundamental task in developmental biology is to
identify the mechanisms which drive morphogenesis. In many cases,
pattern formation is driven by the positional information
determined by both the gradient of maternal factors and hard-wired
mechanisms embedded in the genome. Alternative mechanisms of
positional information that contribute to patterning are the
influence of signals derived from surrounding tissues. In this
paper, we show that the interplay of geometrical constraints
imposed by tissue shapes and hard-wired mechanisms,
computationally implemented by a gene network model, can able to
induce stable complex patterns. The rise of these patterns depends
strongly on the geometrical constraints such as the shape of the
surrounding tissues. \end{abstract}



\maketitle
\section{INTRODUCTION}
Developmental biology has identified several organizing principles
that contribute in an orchestrated manner to embryogenesis and
organogenesis. Several cell communication processes, such as
lateral inhibition \cite{DoeG85}, embryonic induction
\cite{Sand01}, cell growth and death \cite{Adac02} and cell
migration \cite{More98}; guide the development of tissues and
organs. Such the processes are controlled by the information
contained in the genome, which is usually the same in all cells.
Therefore, a question arises: how do complex patterns emerge from
initially homogeneous conditions? One way to address this issue is
by morphogenic gradients. Morphogens are signaling molecules that
can induce several distinct cell fates in a
concentration-dependent manner \cite{Gurd98,Gurd01}. The diffusion
of a morphogen from a localized source generates a graded
distribution across a field of cells. The local concentration of
this molecule would induce the fate of nearby cells in a
position-dependent manner. Typically, a morphogen is produced by a
distinct localized set of cells, from which it moves into
surrounding tissues. If certain conditions are met, a spatial
gradient of morphogen concentration, decaying away from the
source, will result. When a signal is bound to its receptor, a
specific intracellular signal transduction pathway is triggered,
leading to transcriptional and/or pos-transcriptional changes in
responsive cells. If cells can sense the local concentration
accurately, then spatial patterns of cell response can be
generated. For example, in Xenopus embryos, it has been
established that bone morphogenetic protein controls neural and
epidermal fates in the vertebrate ectoderm, under the control of
antagonists secreted by the organizer region of the mesoderm
\cite{Wils95}. Also wingless, hedgehog, and decapentaplegic
proteins in the Drosophila wing imaginal discs, are well studied
examples of secreted signaling proteins \cite{Stri99}.

From a mathematical point of view, Alan Turing showed a pathway to
pattern formation without pre-patterns for its initiation
\cite{Turi52,Koch94}. This self-organized mechanism can be thought
of as a competition between activation by a slow diffusing
chemical and inhibition by a faster chemical, and has been largely
applied to explain biological pattern formation (for example see
\cite{Koch94,Pain99}. However, it is important to keep in mind
that biological strategies of patterning can be very different
from the self-organized Turing-type mechanisms. Segmentation in
Drosophila \cite{Ingh88,Clyd03} and somitogenesis
\cite{Goss98,Saga01} in vertebrate seem to follow a hard-wired
strategy hierarchically organized from small regulatory gene
networks. The extent to which patterns under development are
self-organized or hard-wired remains an issue of debate
\cite{Monk00,Wolp01,Free00}.

In this paper we will explore, by means of computational
simulations, the role of chemical signaling derived from
surrounding tissues to patterning processes. In particular, we
will show that specific geometrical constraints, like the shape of
morphogen sources, are able to generate nontrivial endogenous
gradients which will play a key role in the formation of complex
structures. Modeling pattern formation is a current research topic
in developmental biology \cite{Shva02,Sala02,Dass00,Koch94,Pain99}
and to the best of our knowledge, the interplay of geometrical
constraints, imposed by the shapes of surrounding tissues, and
feedback control implemented in a gene network model has never
been addressed before. Some studies about
the effects of curvature and other geometrical constraint on
pattern formation have been described in \cite{Plaz04,Barr99}.

In most cases, the shape of endogenous morphogen gradients is
rarely known because the absolute concentration of the factor
under analysis is so low. Furthermore, the regulatory gene
networks, as well as the signalling pathway involved in
developmental programs are not, in most cases, quantitatively
characterized. For these reasons, in order to illustrate and to
test the feasibility of our approach, we use a mathematical model
and computational analysis. Our mathematical model sticks to the
hard-wired strategy, implementing several negative feedback loops
in a generic regulatory gene network. This gene network embedded
in endogenous gradients is able to induce stable complex patterns.
The rise of these patterns depends strongly on the geometrical
constraints such as the shape of the surrounding tissues.

Successful development can take place in a range of environments.
Developmental mechanism of patterning must be robust and precise.
Signals produced at wrong place or time can lead to an
inappropriate developmental process. The relevance of noise and
robustness will be also discussed.

\section{METHODS}
\subsection{Shaping morphogen gradients}

Although pattern formation is indeed induced by graded activation
of a signaling pathway, several factors are involved in the
distribution of a morphogen in a developmental field
\cite{Tele01}. The mechanisms involved in gradient shaping is one
of the main issues debated in developmental biology today. Such
mechanisms include passive diffusion in the extracellular matrix
\cite{McDo01}, repeated rounds of endocytosis and exocytosis
\cite{Entc00}, cytonemes \cite{Rami99}, argosomes \cite{Grec01},
and a recently proposed mechanism by mRNA decay in growing
structures \cite{Dubr01}. It is beyond the scope of this paper
modeling the shape gradient genesis. We will adopt the simple
localized source-dispersed sink models, where the shape of a
morphogen gradient is dictated by the rate of diffusion away from
the site of synthesis, along with the rate degradation. The
gradient can be broadened by either increasing the diffusion rate
or decreasing the degradation rate, while the opposite effects can
produce a much steeper gradient. In particular, the diffusive
substances are secreted by the fixed boundary of surrounding
tissues. Thus, we also consider that these substances are degraded
over all developmental field. We can represent these processes by
the following equations
\begin{eqnarray}
\dot{a}&=& D_a \nabla ^2a  -\lambda _a a  \nonumber \\
\dot{h}&=& D_h \nabla ^2h  -\lambda _h h ,
\end{eqnarray}
where $a$ ($h$) denotes the activator (inhibitor) concentration
which is released by the tissue I (tissue II) and degraded over
the domain at rate $\lambda _a$ ($\lambda _h$). Fig. 1 shows a
schematic representation of surrounding tissues, boundaries
(bottom panel) and the concentration profiles $a$ and $h$ over a
horizontal segment in the middle of a semi-elliptic domain (top
panel).

In order to evidence the key role of geometrical constraints in
the patterning process, the overall processes were implemented on
both semi-elliptical and semicircular domains. Fig. 2 depicts the
concentration fields of activators (A) and inhibitors (B) in a
portion of the semi-elliptical domain. For the semicircular domain
the concentration fields of activators and inhibitors are shown in
(A) and (B) respectively and they were obtained using the
parameters in Table 1. Despite of small differences between the
top and bottom morphogen distributions, they are sufficient to
induce completely different patterns of expression of regulated
genes, as we will show later.

\subsection{Hard-wired model for development}
The striped pattern of pair-rule genes in Drosophila is a
consequence of the fact that each stripe is separately controlled
by dedicated transcription regulation \cite{Clyd03}. This is an
example where biological strategies of patterning seem to follow a
hierarchically organized building strategy from small modular
regulatory gene networks. We will construct a small regulatory
network (see Fig. 3) which is able to produce complex patterns
from an initially homogeneous distribution of the genes products.
Our mathematical model considers several biochemical and
biophysical processes currently present in many biological
descriptions, including: diffusion, degradation, noise and of
course, regulatory activation and/or repression of gene
expressions. The last one forms two negative feedback loops, which
seem to be essential control mechanisms in developmental networks
\cite{Free00}. We will describe regulatory activation sigmoidal
functions $s^+\left(x, \theta,n\right)=x^n/(x^n+\theta ^n)$. If
the concentration of activator factor is below the threshold
$\theta$, the gene is poorly expressed, whereas above this
threshold its expression grow until saturate. The regulatory
inhibition will be represented by $s^-=\theta ^n/(x^n+\theta
^n)\equiv 1-s^+$. Sigmoidal functions, in particular Hill
functions, have been used in a variety of genetic regulatory
models \cite{Dass00,Shva02,Rein95,Smol04}. In general, there are
several regulatory interactions acting over a single gene, and in
these cases the resulting expression rates will be described by
the product of sigmoidal functions corresponding to each
interaction \cite{Mest95}, rather than to the additive formulation
used in \cite{Dass00,Shva02,Rein95}. For the sake of simplicity,
our model does not consider other factors as growth, cell
division, pos-transcriptional regulation, etc., and does not
explicitly distinguish between genes and their products.

We will consider five genes which form a small network of
interacting genes. Their product concentrations, represented by
$u_i$ with $i=1,\ldots,5$, evolve following the reaction-diffusion
equations
\begin{equation}
\dot{u_i}= d_{i} \nabla ^2 u_i + f_i \left(a,h,\{u_i\} \right)
-\lambda _{u_i} u_i + \eta \epsilon \left( t \right) u_i,
\end{equation}
where $f_i$ represents the regulatory transcription control of
gene $i$, and $\lambda _i$ is the degradation rate. The last term
represents a multiplicative noise. $\epsilon$ is a random variable
that is assumed uncorrelated and Gaussian distributed (null mean
and variance unitary). $\eta$ represents the intensity of noise.
Eq. (2) governs the spatio-temporal evolution of the small
regulatory gene network represented in Fig. 3. This network
suffers the influence of the surrounding tissues through the
biochemical agents concentration of $a$ and $h$ (which are
constant at a time). The particular form of $f_i$ depends on the
regulatory inputs acting over gene $i$, and will be discussed
later.

In the hierarchically organized hard-wired strategy, the
regulatory role of the genes belonging to a network could be
organized in different levels, for example: positional level,
regulatory level, output level, etc.. These levels interact
between them to constitute a specific network structure. In this
framework, we can understand a complete developmental program as a
network of interacting gene networks.

In this spirit, the positional level will be assigned to U1, the
gene associated to concentration $u_1$. U1 will decode the
positional information derived from the geometrical constraints
(boundaries) through $a$ and $h$ concentrations. The transcription
of gene U1 will be activated when $a$ reaches a threshold $\theta
_{1a}$ and will be inhibited by $h$ and its own product at the
thresholds $\theta _{1h}$ and $\theta _{11}$ respectively (the top
panel of Fig. 1 shows the values of these parameters). Therefore,
the expression of $u_1$ is governed by Eq. 3:
\begin{eqnarray}
f_1\left(a,h,u_1\right) &=&r_{1} \ s^+\left(a,\theta
_{1a},n_{1a}\right) s^-\left(h,\theta _{1h},n_{1h}\right) \times
\nonumber \\ &&s^-\left(u_1,\theta _{11},n_{11}\right).
\end{eqnarray}

The expression of U1 activates in a concentration-dependent manner
the genes U2, U3, and U4, (associated to the concentrations $u_2,
\ u_3$ and $u_4$ respectively), which constitutes essentially a
regulatory level of the hierarchy. U2, U3, and U4 are activated at
thresholds $\theta _{21} > \theta _{31} > \theta _{41}$,
respectively, and each product inhibits the transcription of the
other genes, thus forming two negative feedback loops. So, we can
write:
\begin{eqnarray}
f_2\left(\{u_i\}\right)&=&r_{22} \ s^+\left(u_2,\theta
_{22},n_{22}\right) + r_{21} \ s^+\left(u_1,\theta
_{21},n_{21}\right) \times \nonumber \\  &&s^-\left(u_3,\theta
_{23},n_{23}\right)s^-\left(u_4,\theta _{24},n_{24}\right),
\end{eqnarray}
where $\{u_i\}$ represents the set of variables
$\{u_1,u_2,u_3,u_4,u_5\}$. for $r_{22}>0$ this gene is also self
activated. We have added a self activated term in (4) to study
their effects over noise control.
\begin{eqnarray}
f_3\left(\{u_i\}\right)&=&r_{3} \ s^+\left(u_1,\theta
_{31},n_{31}\right)  s^-\left(u_2,\theta _{32},n_{32}\right)
\times \nonumber \\ &&s^-\left(u_4,\theta _{34},n_{34}\right),
\end{eqnarray}
\begin{eqnarray}
f_4\left(\{u_i\}\right)&=&r_{4} \ s^+\left(u_1,\theta
_{41},n_{41}\right) s^-\left(u_2,\theta _{42},n_{42}\right) \times
\nonumber \\ &&s^-\left(u_3,\theta _{43},n_{43}\right).
\end{eqnarray}

Thus the gradient associated to $u_1$ generates discrete spatial
domains of regulatory factors which interact between them and
generate a more complicated spatial domain of transcription of the
third level of the hierarchy, the output level. For simplicity, we
consider also that this level is constituted by only one gene. Its
activation is promoted by $u_1$ and it is down-regulated by the
gene products of the former level, $u_2, \ u_3$ and $u_4$.
\begin{eqnarray}
f_5\left(\{u_i\}\right)&=&r_{5} \ s^+\left(u_1,\theta
_{51},n_{51}\right) s^-\left(u_2,\theta _{52},n_{52}\right) \times
\nonumber \\ &&s^-\left(u_3,\theta
_{53},n_{53}\right)s^-\left(u_4,\theta _{54},n_{54}\right).
\end{eqnarray}

The positive and negative feedback loops of the model are shown in
Fig. 3. It is important to keep in mind that the model was
constructed to illustrate that gene networks, embedded in feedback
control principles, can induce or not complex patterns depending
on geometrical constraints of the problem. For the aim of our
study, this small network is enough. Of course, other kinds of
network can be designed to model specific developmental systems.

\section{RESULTS}
\subsection{Geometrical constraints and gene regulatory network}
The boundary problem is completely specified by Eqs. (1), the
initial condition (IC) and the boundary condition (BC). In this
sense, $a=1$ over $C$ and null over $C'$, the two vertical
boundary segments are connected by a periodical boundary
condition, while the inhibitor is $h=1$ over $C'$ and null over
$C$, the two vertical boundary segments are connected by periodic
boundary condition. Initially, both $a$ and $h$ concentrations are
null over all domain compatible with BC. We integrate numerically
the Eqs. (1) for two different situations which differ in
degradation parameter values. The simulations run for a long
period of time (5,000 steps) after which the resulting
concentration $a$ and $h$ are stationary. Then, these
concentrations will be used as input signals acting over the
network in several different conditions.

The aim of the paper is to study the effects of geometrical
constraints on the patterning in developmental motivated models
rather than to modeling a particular system. For this reason it is
not necessary to fit the parameters to reproduce experimental
observation. However, as the particular model chosen for our
purpose has 52 parameters, we need to establish some guidelines so
as to reach stable stationary patterns before implementing massive
computation. In this model, the gene expression rates involve Hill
functions whose coefficients $n_{i,j}$ range from 1 to 5.
$n_{i,j}>1$ reflects the fact that {\it cis}-regulatory systems
have usually many binding sites for each {\it trans}-regulatory
element. These parameters can influence the shape patterns, in
particular by controling the sharpness of the patterns. The
$\theta_{i,j}$ parameters represent the threshold for switching on
or off the transcription processes. They were setted in the
interval [0.0,1.2]. Some patterns are very sensitive to these
parameters while others are robust. In particular, the thresholds
corresponding to the activation of U2, U3, and U4 by U1 are sorted
in decreasing order so as to form a sequence of activation; we
setted them to $\theta _{41}=0.4, \  \theta _{31}=0.6$ and $
\theta _{21}$ ranges from 0.9 to 1.2. The diffusion constants and
degradation rates are initially the same for all biochemical
species (0.01 and 0.20 respectively). Initially, the level of
noise was null for all species. Table 1 and 2, display the
parameters values used for most computations, otherwise we will
mention other parameters values.

The set of partial differential equations were discretized using
standard finite difference methods. The resulting large-scale
dynamical system was integrated in time using forward Euler
integration. We have performed simulations for different sets of
parameters subject to the above mentioned restrictions. We have
found that patterns are sensitive to the spatial profile
concentration of $a$ and $h$ and the parameters that regulate the
activation and repression of U1 expression, $\theta _{1a}$,
$\theta _{1h}$ and $\theta _{11}$. For any given profiles $a$ and
$h$, there are regions in the parameters space where no patterns
form, mainly due to the fact that activator concentration $a$ does
not reach the threshold $\theta _{1a}$. An interplay of these
parameters and both $a$ and $h$ profiles control the position and
the coarse-grain aspect of the pattern shapes.

Figure 4 depicts the density of concentrations of $u_2$ (A), $u_3$
(B), $u_4$ (C) and $u_5$ (D) obtained using the input profile 1
(see Table 1) after 400 time steps, when stationarity is reached.
We can see that the concentrations present complementary aspects.
In general, the pattern of the output level is particularly
complex as it is under the effects of four regulatory fields. In
contrast, the concentration corresponding to $u_2$ is the simplest
one because the regulatory field is almost homogenous in its
expression domain. We have observed trough several simulations
that these characteristics are quite general.

The remarkable observation in our study is the fact that
geometrical constraints play a key role when surrounding tissues
secret morphogenic substances. Localized patterns, similar to
those developed in the semi-elliptic domain, were not formed in a
semi-circular domain. In the last case, we were not able to tune
the parameters in order to obtain patterns different from
semi-circular stripes as shown in Figure 5. The reason for that is
based on the fact that (despite of diffusively substance $a$ and
$h$ are uniformly secreted by the surrounding tissue) the
curvature of the tissues is not homogenous in the elliptic case,
but it is homogenous in the circular case. The two fronts of
secreted substances are able to generate a localized expression
domain for U1 in the non-homogenous curvature case. The position
where the complex structures develop in Figure 4 corresponds to
the maximal value of the surrounding tissue curvature. In
contrast, for homogenous curvature the expression domains are not
localized and that does not depend on tuning adequately the
network parameters.

The same network can generate different patterns depending on the
parameters values and the input profiles. Fig. 6 illustrates four
patterns of expression level corresponding to $u_5$ after 400
steps. The patterns displayed at the top (A and B panels) have
been obtained using profiles 1 (see Table 1) and two different
sets of parameter values. In panel A the system reached
stationarity, while panel B reached stationarity after 800 steps
(see Movie 1). Stationary patterns displayed at the bottom (C and
D panels) have been obtained using profiles 2 (see Table 1). The
parameter values are shown in Table 3. Often different sets of
parameter values lead to the same expression pattern, but the way
to reach this pattern can differ from set to set. The movie 1
illustrates the evolution of the network in two different
conditions. The right panel of the movie corresponds to the same
condition as in Fig. 6B, while in the left panel we changed
slightly the value of parameter to $r_3=32.1 \times 10^{-2}$. Both
simulations correspond to 800 steps. We can see that the final
patterns are almost the same, but their evolutions are quite
different. Small perturbations of boundary conditions affect the
patterning processes differently, depending on the network
parameters. For example, Fig. 7 shows patterns generated by the
network in the same conditions as in Fig. 6, but with a
perturbation in the boundary tissues (both in $C$ and $C'$). The
panels of Fig. 7 illustrate that the perturbation can affect a
little, as in panel A, or too much, destroying completely the
patterns as in panels C and D.  This result suggests that the
shape of the tissue strongly influences the formation of pattern
expression and biochemical interactions are not a necessary
condition to explain alterations in the resulting gene expression
patterns. The influence could be through the geometrical
constraint imposed by the surrounding tissues.

In general, genetic network which underlie the developmental
programs of living cells, must withstand considerable random
perturbations. This occurs as fluctuation in, for example,
transcription, translation, and RNA and protein degradation
\cite{Arki98,McAd97}. We have also studied by numerical simulation
the effects of the noise on the patterns in two situations. For
all  the previous cases we have setted $r_{22}=0.0$, i.e. there is
no autocatalytic feedback acting over $u_2$. It would be
interesting to compare the effect of noise when this positive
feedback loop is present in the network. Figures 8 and 9 show the
expression profiles of $u_3$ and $u_5$ respectively, obtained for
five different levels of noise $\eta=0.00$ (A and F), $\eta=0.05$
(B and G), $\eta=0.10$ (C and H), $\eta=0.15$ (D and I) and
$\eta=20$ (E and J). In both figures the top panels (A-E)
correspond to the network without the above mentioned
autocatalytic loop, while the bottom panels correspond to the
network with the loop. The network parameter values used in both
cases were the same as those used in Figure 7A, except for
parameters $r_{21}$ and $r_{22}$, which in the case of the network
with the loop, change from $6.85 \times 10^{-2}$ to $4.84 \times
10^{-2}$, and from zero to $4 \times 10^{-2}$ respectively. We can
see that pattern structure remains almost unaltered for noise
level smaller than $\sim 10\%$ for the network with the
autocatalytic loop. On the other hand, when this loop is absent
($r_{22}=0.0$), the patterns seem to be more noise sensitive. This
example illustrate that the network is robust against a
considerably high level of noise. This robustness could arise from
the fact that there are two negative loops underlying the
regulatory interactions. Furthermore, this example shows, as in
many homeotic genes, that the positive feedback loop helps the
maintenance of their expression, thus contributing to the
stabilization of the patterns.

\subsection{Geometrical constraints and self-organized strategy}
In order to study the influence of geometrical constraints on a
self-organized system type, we have also implemented the
Gray-Scott model on the semi-circular and semi-elliptical domains.
The Gray-Scott model is a variant of the autocatalytic Selkov
model of glycolysis which corresponds to the following two
irreversible reactions:
\begin{eqnarray}
U \ + \ 2V \ &\rightarrow & \ 3V \nonumber  \\
V \ &\rightarrow & \ P. \nonumber
\end{eqnarray}
P is an inert product. Both U and V are removed by the feeding
process. A variety of spatio-temporal patterns were derived in
response to finite-amplitude perturbation \cite{Pear93}. We
consider that $u$ and $v$ are coupled to the activator field $a$
and an to inhibitor $h$ as follows:
\begin{eqnarray}
\dot{u}&=& D_{u} \nabla ^2 u - uv^2 + F(1-u) + r_u \ f\left(a,h\right) \nonumber \\
\dot{v}&=& D_{v} \nabla ^2 v + uv^2 - (F+k)v + r_v \
g\left(a,h\right),
\end{eqnarray}
where
\begin{eqnarray}
f\left(a,h\right) &=& s^+\left(a,\theta_{u,a},n_{u,a}\right)s^-\left(h,\theta_{u,h},n_{u,h}\right) \nonumber \\
g\left(a,h\right) &=&
s^+\left(a,\theta_{v,a},n_{v,a}\right)s^-\left(h,\theta_{v,h},n_{v,h}\right).
\nonumber
\end{eqnarray}

We study numerically the response to different positioned
perturbations of the system which evolve in both geometries
mentioned above. The initial condition corresponds to the trivial
state ($u=1$ and $v=0$). A 16 by 16 mesh point area located over
the semimajor (horizontal) axis was perturbed to $u=0.50$ and
$v=0.25$. The system suffers the influence of $a$ and $h$ fields
determined by the profile 1 (see Table 1). Figure 10 shows the
patterns obtained from the system described by Eqs. (8) on the
elliptic geometry (for system response on the circular geometry
see Fig. 11). The horizontal position of perturbations were
centered at 38.3, 42.3, 46.3, 50.3 and 54.3; while parameter $k$
ranges from 0.060 to 0.066 and is uniformly spaced. Other adopted
parameters are shown in Table 4. As we can see, for the several
values of the parameter $k$, the final patterns upon 2,000 time
steps depend strongly on the position where the perturbation
occurs; while the geometry of the domain, either elliptical or
circular, seems to have very little influence. It should be
mentioned that, in contrast to the patterns derived in the
previous section, these patterns are not stable, they continue
growing until all the domain is filled.

\section{DISCUSSION}

Early development of multicellular organisms is marked by a rapid
initial increase in their cell numbers, accompanied by
morphogenetic processes leading to the gradual formation of organs
of characteristic shapes. During morphogenesis, through
differentiation under strict genetic control, cells become more
and more specialized. Further, genetic mechanisms as morphogenesis
also require generic physical principles, such as, diffusion,
spreading, differential adhesion, chemotaxis, etc.. As a
consequence, development rely on an intricate interplay of generic
and genetic mechanisms. In this paper, we have addressed the issue
on how chemical signaling derived from surrounding tissues can
drive patterning processes. Our results, derived by computer
simulation, suggest that the shape of the source (tissue) and
other geometrical constraints strongly influence the formation of
complex structures. In particular, we found that non-homogenous
curvature of signaling tissues can generate complex patterns when
acting on gene network, while the same network embedded in a
geometry with homogenous curvature (boxes and circles) does not
form localized structures. This result suggests an important
consequence for development: let us consider that the shape of an
organizer (tissue I or II in our case) is controlled by a gene
network A, and the genic response to the organization field is
controlled by a network B (the network shown in Fig. 3 in our
case), which is not regulated by genes of network A. Any mutation
in A that alters the shape of the organizer and consequently the
organization field, will affect the development of the induced
structure or organ. Thus, we can conclude that biochemical
interaction is not a necessary condition to explain alteration in
the overall induced output; the interaction could be underlaid by
a geometrical constraint imposed by the organizer. In contrast to
this hard-wire system, the self-organized system examined here
(Gray-Scott model) seems to be more insensitive to the gradients
derived from surrounding tissues.

We have also studied the robustness of patterns in
response to noise. In this sense, two kinds of network have been
used to illustrate the effect of noise in either the presence or
absence of a positive feedback loop. The pattern driven by the
first network seems to present higher robustness to noise than the
second one. However, the two negative feedback loops presented in
both cases were able to guarantee the stability of the pattern
structure against a considerably high level of noise.

Although we have used a particular network topology with several
mutually-inhibiting factors, there are many other
possibilities for the topology that could produce interesting
patterns. Each topology can response in a different way to the
intrinsic noise of the patterning process. The particular case
examined here was enough to conclude that the hard wired
mechanism, as the regulatory networks can create complex
biological shapes out of a simple structure. This mechanism could
be enriched by incorporating geometrical constraint as key
ingredients of morphogenesis processes.

\section*{Acknowledgments}
Luciano da F. Costa is grateful to FAPESP (process 99/12765-2),
CNPq (process 308231/03-1) and Human Frontier (RGP 39/2002) for
financial support. Luis Diambra thanks Human Frontier for his
post-doc grant.

\newpage

\section*{Tables}

\begin{table}[ht]
\caption{Parameter values for diffusion, degradation and
geometrical constraints used to obtain the input profiles 1 and 2.
These profiles differ in the degradation values $\gamma_{a}$ and
$\gamma_{h}$ which were reduced to $0.5\times 10^{-3}$ to generate
the input profile 2.}
\begin{tabular}{c|l|c|l}
\multicolumn{4}{c}{}\\
  Parameters & Values   &  Parameters & Values    \\
  \hline
$d_{i}$        &\ 0.01                 & \underline{Semi-ellipse}   &          \\
$\gamma_{i} \ (i\neq 5$)  &\ 0.20      & internal semimajor axis  &\ 40.00    \\
$\gamma_{5}$   &\ 0.35                 & internal semiminor axis  &\ 14.14     \\
$D_{a}$        &\ 0.04                 & external semimajor axis  &\ 86.60    \\
$D_{h}$        &\ 0.04                 & external semiminor axis  &\ 31.62    \\
$\gamma_{a}$   &\ $1.0 \times 10^{-3}$ & \underline{Semi-circular}  &          \\
$\gamma_{h}$   &\ $1.0 \times 10^{-3}$ & internal radius       &\ 24.70     \\
$\Delta t$     &\ 1.00                 & external radius       &\ 51.96    \\
$\Delta x$     &\ 0.40                 &  \\
\hline
\end{tabular}
\end{table}

\newpage

\begin{table}[ht]
\caption{Parameter values of the gene network used to obtain the
pattern shown in Figure 4. These values were also used in other
simulations whenever we have not stated otherwise (see also Table
III).}
\begin{tabular}{c|l|c|l}
\multicolumn{4}{c}{}\\
  Parameters & Values  & Parameters & Values  \\
  \hline
$r_{1}$        &\ 6.25                   &  $r_{4}$        &\ $5.8 \times 10^{-2}$ \\
$\theta_{1a}$  &\ 0.40                   &  $\theta_{41}$  &\ 0.40                  \\
$\theta_{1h}$  &\ 0.05                   &  $\theta_{42}$  &\ 0.25                  \\
$\theta_{11}$  &\ 0.20                   &  $\theta_{43}$  &\ 0.20                  \\
$n_{1a}$       &\ 1.00                   &  $n_{41}$       &\ 4.00                  \\
$n_{1h}$       &\ 3.00                   &  $n_{42}$       &\ 2.00                  \\
$n_{11}$       &\ 5.00                   &  $n_{43}$       &\ 2.00
\\ \hline
$r_{21}$       &\ $8.06 \times 10^{-2}$  &  $r_{5}$        &\ 20/27                 \\
$\theta_{21}$  &\ 1.05                   &  $\theta_{51}$  &\ 0.55                  \\
$\theta_{23}$  &\ 0.35                   &  $\theta_{52}$  &\ 0.20                  \\
$\theta_{24}$  &\ 0.45                   &  $\theta_{53}$  &\ 0.30                  \\
$n_{21}$       &\ 4.00                   &  $\theta_{54}$  &\ 0.50                  \\
$n_{23}$       &\ 2.00                   &  $n_{51}$       &\ 4.00                  \\
$n_{24}$       &\ 2.00                   &  $n_{52}$       &\ 3.00
\\ \cline{1-2}
$r_{3}$        &\ 0.21                   &  $n_{53}$       &\ 3.00                  \\
$\theta_{31}$  &\ 0.60                   &  $n_{54}$       &\ 3.00
\\ \cline{3-4} $\theta_{32}$  &\ 0.10                   &
\multicolumn{2}{c}{PFL parameters}      \\ \cline{3-4}
$\theta_{34}$  &\ 0.10                   & $r_{22}$        &\ 0.00                  \\
$n_{31}$       &\ 4.00                   & $\theta_{22}$   &\ 0.10                  \\
$n_{32}$       &\ 2.00                   & $n_{22}$        &\ 3.00                  \\
$n_{34}$       &\ 2.00                   &                 &                        \\
\hline
\end{tabular}
\end{table}
\newpage
\begin{table}[ht]
\caption{Parameter values of the gene network used to obtain the
patterns shown in Figure 7. The values of parameters not shown in
this table are the same as in Table II.}
\begin{tabular}{c|l|c|l|c|l|c|l}
\multicolumn{8}{c}{}\\
\hline
\multicolumn{2}{c}{Fig. 7A} & \multicolumn{2}{|c}{Fig. 7B} &
\multicolumn{2}{c}{Fig. 7C} & \multicolumn{2}{|c}{Fig. 7D} \\

\hline Parameters & Values  & Parameters & Values  & Parameters & Values  & Parameters & Values  \\  \hline
$r_{21}$       &\ $6.85  \times 10^{-2}$   &\ $r_{1}$        &\ 9.25                   &$r_{1}$        &\ 7.50                     &\ $r_{1}$        &\ $7.50 \times 10^{-6}$ \\
$\theta_{21}$  &\ 0.99                     &\ $\theta_{11}$  &\ 0.265                  &$\theta_{1a}$  &\ 0.20                     &\ $r_{21}$       &\ $8.46 \times 10^{-2}$ \\
$n_{21}$       &\ 5.00                     &\ $r_{21}$       &\ $12.09 \times 10^{-2}$ &$r_{21}$       &\ $12.09 \times 10^{-2}$   &\ $\theta_{21}$  &\ 0.95                  \\
$r_{3}$        &\ 0.215                    &\ $\theta_{21}$  &\ 0.85                   &$\theta_{21}$  &\ 1.30                     &\ $n_{21}$       &\ 5.00                  \\
$\theta_{31}$  &\ 0.65                     &\ $n_{21}$       &\ 5.00                   &$n_{21}$       &\ 5.00                     &\ $r_{3}$        &\ 0.27                  \\
$r_{4}$        &\ $5.60 \times 10^{-2}$    &\ $r_{3}$        &\ $32.02 \times 10^{-2}$ &$r_{3}$        &\ 0.18                     &\ $n_{31}$       &\ 2.00                  \\
$r_{5}$        &\ 0.18                     &\ $r_{4}$        &\ $6.24 \times 10^{-2}$  &$r_{4}$        &\ $5.20 \times 10^{-2}$    &\ $r_{4}$        &\ $9.75 \times 10^{-2}$ \\
$\theta_{51}$  &\ 0.50                     &\ $r_{5}$        &\ 35/216                 &$r_{5}$        &\ 10/9                     &\ $\theta_{41}$  &\ 0.45                  \\
$\theta_{52}$  &\ 0.55                     &\ $\theta_{51}$  &\ 0.50                   &$\theta_{51}$  &\ 0.52                     &\ $\theta_{42}$  &\ 0.20                  \\
$\theta_{53}$  &\ 0.50                     &\ $\theta_{52}$  &\ 0.40                   &$\theta_{52}$  &\ 0.40                     &\ $n_{41}$       &\ 2.00                  \\
$\theta_{54}$  &\ 0.20                     &\ $n_{51}$       &\ 2.00                   &$\theta_{53}$  &\ 0.30                     &\ $r_{5}$        &\ 8.54                  \\
$n_{51}$       &\ 2.00                     &                 &                         &$\theta_{54}$  &\ 0.25                     &\ $\theta_{51}$  &\ 0.50                  \\
               &                           &                 &                         &   $n_{51}$    &\ 3.00                     &\ $\theta_{52}$  &\ 0.40                  \\
               &                           &                 &                         &               &                           &\ $\theta_{53}$  &\ 0.20                  \\
               &                           &                 &                         &               &                           &\ $\theta_{54}$  &\ 0.20                  \\
               &                           &                 &                         &               &                           &\ $n_{51}$       &\ 2.00                  \\
\hline
\end{tabular}
\end{table}
\newpage
\begin{table}[ht]
\caption{Parameter values for the Gray-Scott model.}
\begin{tabular}{c|l|c|l}
\multicolumn{4}{c}{}\\
  Parameters & Values   &  Parameters & Values    \\
  \hline
$F$        &\ 0.04      & $\theta_{ua}(=\theta_{uh})$ & 0.50    \\
$r_{u}$    &\ 0.001     & $\theta_{va}=\theta_{vh})$ & 0.50    \\
$r_{v}$    &\ 0.002     & $n_{ua}(=n_{uh})$ &\ 1.0     \\
$D_{u}$    &\ 0.0336    & $n_{va}(=n_{vh})$ &\ 1.0    \\
$D_{v}$    &\ 0.0168    &  & \\
\hline
\end{tabular}
\end{table}

\newpage

\section*{Figures and movie}
\begin{figure}[h]
\includegraphics[angle=0,scale=0.45]{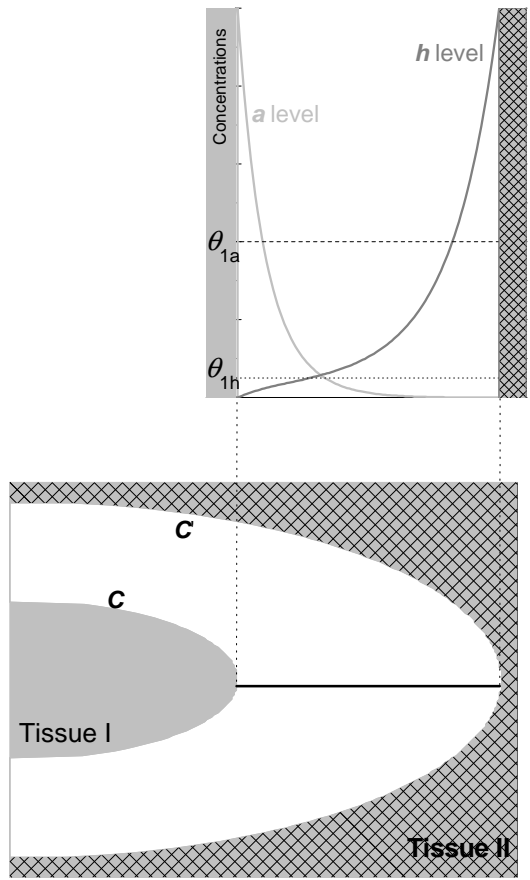}
\caption{A schematic representation of surrounding tissues,
boundaries (bottom) and the concentration profile of activator $a$
and inhibitor $h$ over a horizontal segment in the middle of a
semi-elliptic domain (top).}
\end{figure}
\newpage
\begin{figure}[h]
\includegraphics[angle=0,scale=0.65]{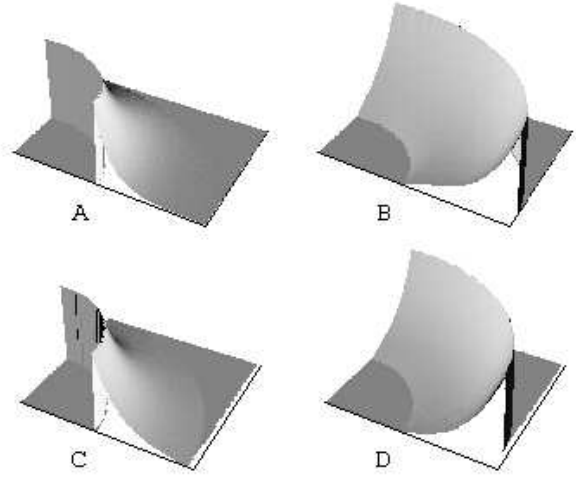}
\caption{3D plots of the concentration of $a$, secreted by the
tissue I (A and C), and $h$ (B and D) which is secreted by the
tissue II, obtained from Eqs. 1, after 5,000 time steps. The
panels A and B correspond to the elliptical geometry, and panels C
and D correspond to the circular geometry.}
\end{figure}
\newpage
\begin{figure}[h]
\includegraphics[angle=0,scale=0.55]{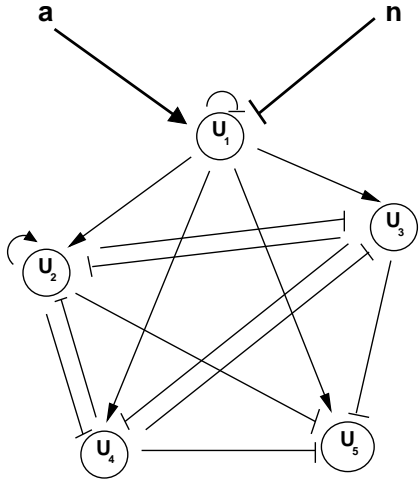}
\caption{Schematic diagram of the network. Circles indicate genes,
blunt-end arrows denote inhibition and pointed arrows denote
activation. Regulatory activation is represented by arrows and
regulatory inhibition by small perpendicular edges, as usually.
This network presents two negative feedback loops, one where U2
inhibits U3 that inhibits U4 that inhibits U2; and another which
is formed by U4 that inhibits U3 that inhibits U2 that inhibits
U4. There is also an auto-inhibitory loop acting over U1 and an
autocatalytic loop acting over U2.}
\end{figure}
\newpage
\begin{figure}[h]
\includegraphics[angle=0,scale=0.70]{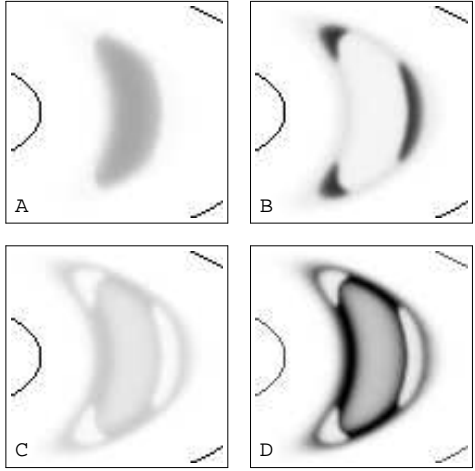}
\caption{Density plot of concentrations of $u_2$ (A), $u_3$ (B),
$u_4$ (C) and $u_5$ (D) after 400 steps in a semi-elliptic domain.
The model parameters used to compute these patterns are displayed
in Table.}
\end{figure}
\newpage
\begin{figure}[h]
\includegraphics[angle=0,scale=0.70]{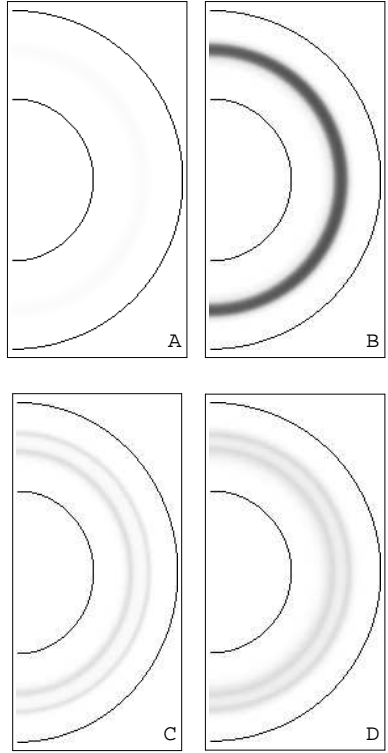}
\caption{Density plot of concentrations of $u_2$ (A), $u_3$ (B),
$u_4$ (C) and $u_5$ (D) after 400 time steps. In this
semi-circular domain, no localized structure were observed for any
set of parameter values.}
\end{figure}
\newpage
\begin{figure}[h]
\includegraphics[angle=0,scale=0.70]{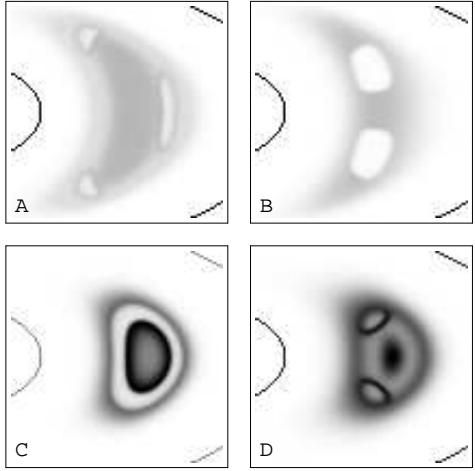}
\caption{Density plot of concentrations of $u_5$ after 400
obtained for inputs profile 1 (A and B) and inputs profile 2(C and
D) with the network parameters displayed in Table 3.}
\end{figure}
\newpage
\begin{figure}[h]
\includegraphics[angle=0,scale=0.70]{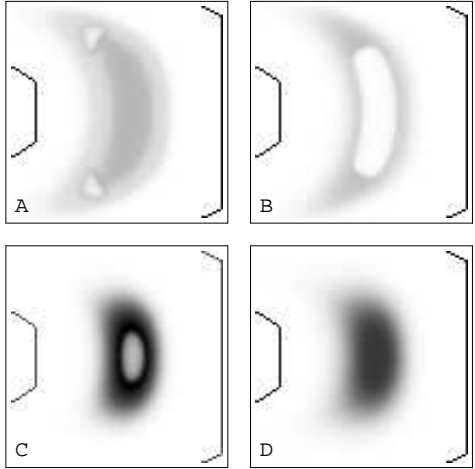}
\caption{Density plot of concentrations of $u_5$ after 400 steps
obtained in the same condition that Fig. 6 with the exception in
the boundary condition which were perturbed.}
\end{figure}
\newpage
\begin{figure}[h]
\includegraphics[angle=0,scale=0.95]{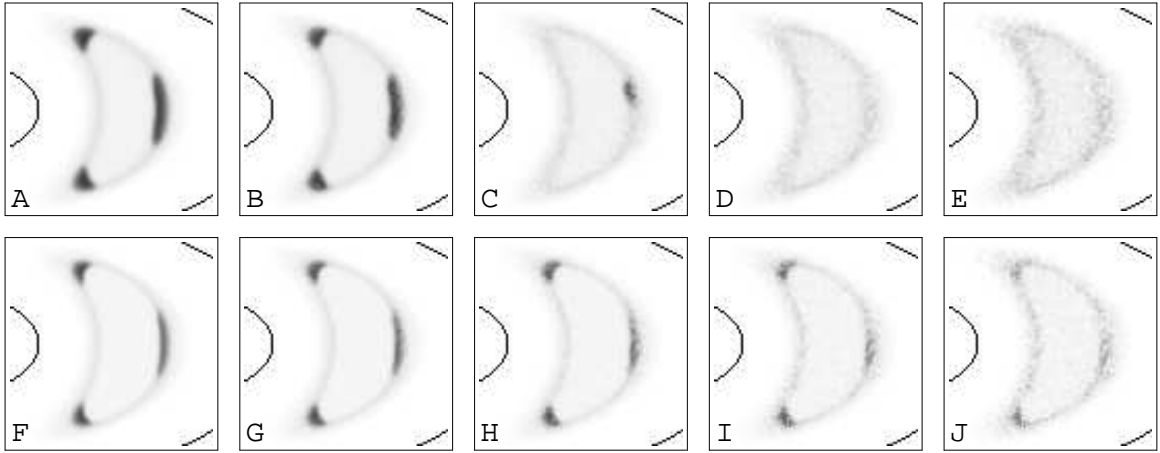}
\caption{Density plot of $u_3$ obtained, after 400 steps, when the
dynamics of the gene network is perturbed with noise. The top
panels correspond to patterns obtained from the network without
the positive feedback loop for increasing level of noise (sorted
from left to right). The bottom panels correspond to patterns
obtained from the network with the feedback loop for increasing
level of noise (sorted from left to right). The level of noise
used were $\eta=0.0$ (A and F) $\eta=0.05$ (B and G), $\eta=0.10$
(C and H), $\eta=0.15$ (D and I), and $\eta=0.20$ (E and J). For
further information, see the text.}
\end{figure}
\newpage
\begin{figure}[h]
\includegraphics[angle=0,scale=0.95]{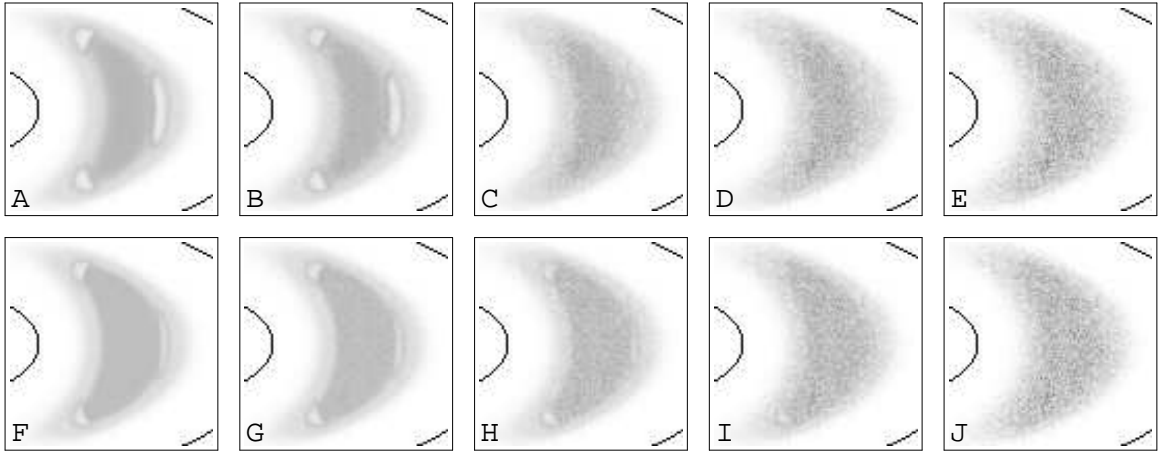}
\caption{Density plot corresponding to $u_5$ obtained in the same
situation as in Fig. 8.}
\end{figure}
\newpage
\begin{figure}[h]
\includegraphics[angle=0,scale=1.0]{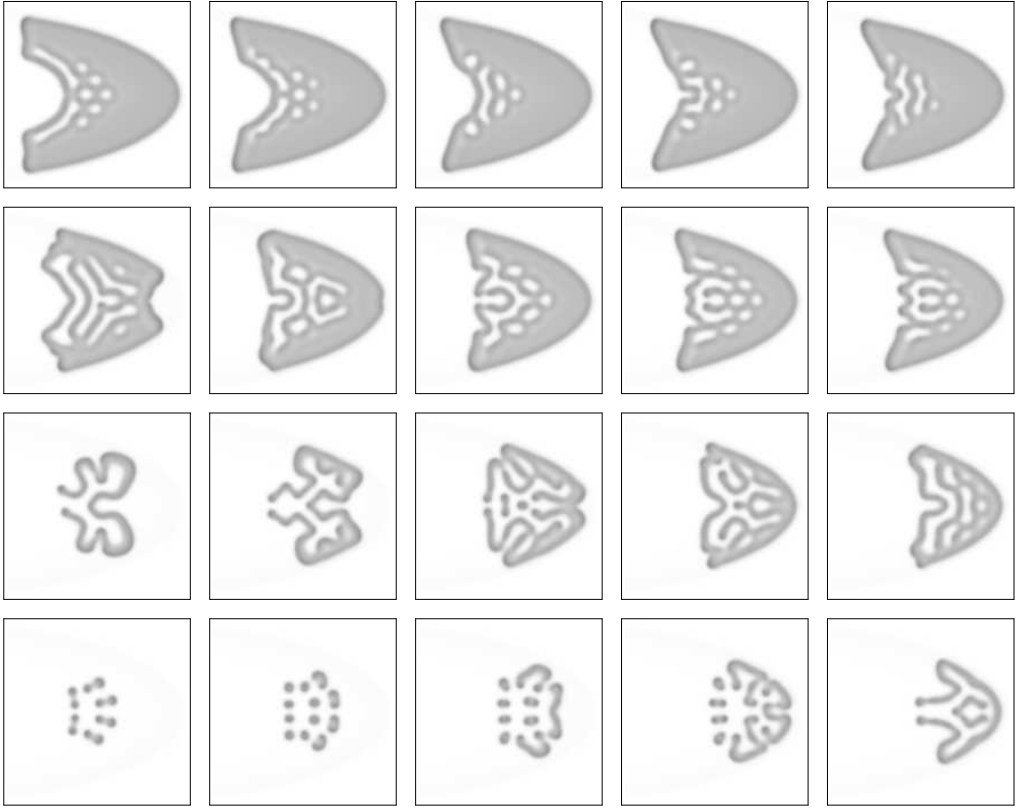}
\caption{Elliptic Geometry: Density plot of the $v$ concentration
for different position of perturbation, increasing from left to
right, and for different values of parameter $k$ which increases
from top to bottom (see details in text).}
\end{figure}
\newpage
\begin{figure}[h]
\includegraphics[angle=0,scale=0.75]{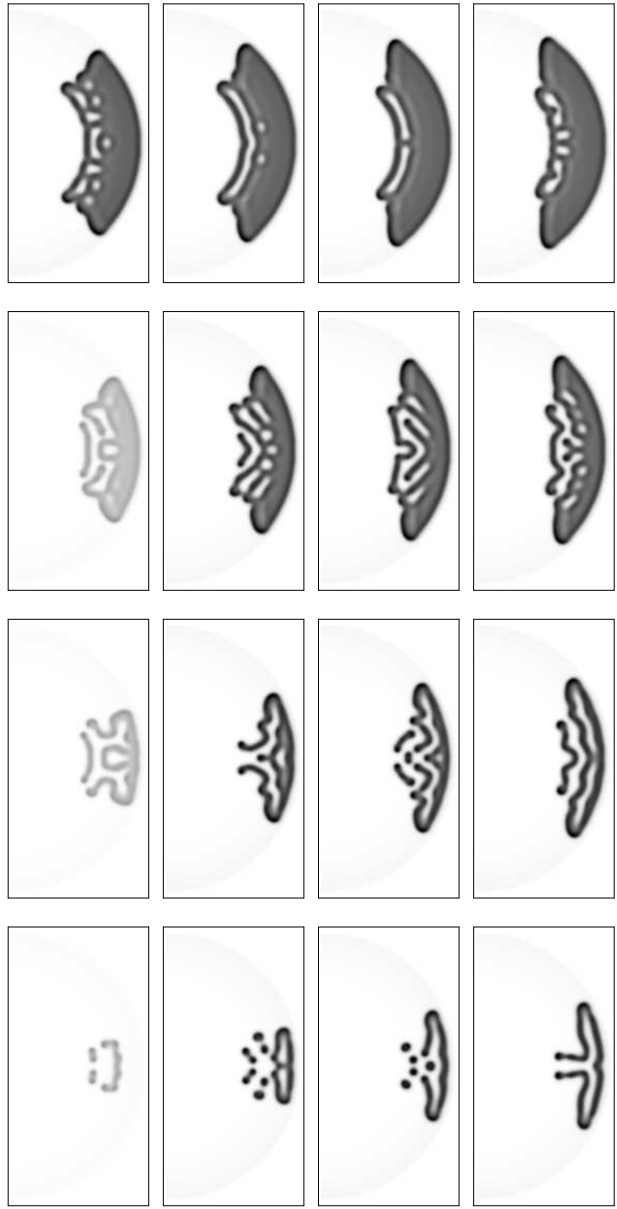}
\caption{Circular Geometry: Density plot of the $v$ concentration
for different positions of perturbation (increasing from left to
right) and for different values of parameter $k$ (increasing from
top to bottom).}
\end{figure}

\newpage
\begin{center}
\begin{picture}(275,200)(0,50)
\put(-20,20){\framebox(275,200)[lb]}
\end{picture}
\end{center}
\vspace*{2cm} Movie 1: (Please see at
http://glia.if.sc.usp.br/luis/teste.mov) The left panel
corresponds to the temporal evolution of the pattern Fig. 7B
during 800 steps. The right panel corresponds to the evolution in
the same condition with the exception of pattern $r_3$ which
changes from $32.02 \times 10^{-2}$ to $32.1 \times 10^{-2}$

\end{document}